# Superconductivity in undoped CaFe$_2$As$_2$ single crystals

Dong-Yun Chen[1], Jia Yu[1], Bin-Bin Ruan[1], Qi Guo[1], Lei Zhang[2], Qing-Ge Mu[1], Xiao-Chuan Wang[1], Bo-Jin Pan[1], Gen-Fu Chen[1,3], Zhi-An Ren[1,3]*

[1] Institute of Physics and Beijing National Laboratory for Condensed Matter Physics, Chinese Academy of Sciences, Beijing 100190, China

[2] High Magnetic Field Laboratory, Chinese Academy of Sciences, Hefei 230031, China

[3] Collaborative Innovation Center of Quantum Matter, Beijing 100190, China

* Email: renzhian@iphy.ac.cn

## Abstract

Single crystals of undoped CaFe$_2$As$_2$ were grown by a FeAs self-flux method, and the crystals were quenched in ice-water rapidly after high temperature growth. The quenched crystal undergoes a collapsed tetragonal structural phase transition around 80 K revealed by the temperature dependent X-ray diffraction measurements. Superconductivity below 25 K was observed in the collapsed phase by resistivity and magnetization measurements. The isothermal magnetization curve measured at 2 K indicates that this is a typical type-II superconductor. For comparison, we systematically characterized the properties of the furnace cooled, quenched, and post-annealed single crystals, and found strong internal crystallographic strain existing in the quenched samples, which is the key for the occurrence of superconductivity in the undoped CaFe$_2$As$_2$ single crystals.

# 1. Introduction

For the recently discovered iron-based high-$T_c$ superconductors [1], superconductivity was usually induced by suppressing the low temperature antiferromagnetic (AFM) transition via chemical doping, or applying external pressure in the parent compounds [2, 3]. Due to the feasible high quality singe crystal growth by flux method for the ternary 122-type $AFe_2As_2$ (A= Ba, Sr, Ca, Eu) superconductors with the $ThCr_2Si_2$-type crystal structure [4, 5], these systems have been widely studied for their physical properties and high-$T_c$ mechanism [6]. Interestingly, superconductivity was also reported in the undoped $SrFe_2As_2$ ($T_c$ up to 21 K) and $BaFe_2As_2$ ($T_c$ up to 22.5 K) single crystals at ambient pressure without any chemical doping, which was suggested to be induced by internal crystallographic strain or self-doping by crystal defects [7, 8].

Among these 122-type iron pnictides, $CaFe_2As_2$ is very particular, which exhibits extreme sensitivity to external pressure on its crystal structure and magnetic properties at ground state [9-15], and the rare-earth doped $Ca_{1-x}Pr_xFe_2As_2$ even reaches an unusual high-$T_c$ of 49 K with the mechanism still unclear [16, 17]. For the undoped $CaFe_2As_2$ compound, it generally crystallizes in the tetragonal $ThCr_2Si_2$-type crystal structure at room temperature, and undergoes a first-order phase transition into an orthorhombic phase at 170 K [18-20], which is also accompanied by a commensurate AFM transition [21]. After applying a pressure above 0.3 GPa, a collapsed, non-magnetic tetragonal phase instead of the AFM phase is formed at low temperature [9, 11-14, 22], and this collapsed phase has a distinct c-axis shrinkage due to the promotion of the inter-layer As-As bonds [23-25]. This collapsed phase was also reported to be stabilized by annealing and quenching processes without applying external pressure [26, 27]. In addition, chemical doping was also reported to be able to stabilize this collapsed phase at ambient pressure [28-31]. Superconductivity up to 12 K was ever reported in undoped $CaFe_2As_2$ crystal under near-hydrostatic pressures [9-11], but after using helium gas as a hydrostatic pressure medium, the signal of superconductivity became absent [13, 14].

Here we report the superconductivity with $T_c \sim 25$ K at ambient pressure in undoped CaFe$_2$As$_2$ single crystals processed by a rapid quenching treatment after high temperature growth. More interestingly, this high-$T_c$ superconductivity is originated in the collapsed tetragonal CaFe$_2$As$_2$ phase.

## 2. Experimental details

High quality single-crystalline samples of CaFe$_2$As$_2$ were grown in a FeAs self-flux method. The FeAs precursor was pre-sintered by solid-state reaction of Fe powder and As powder at 750 $^\circ$C for 30 h in an evacuated quartz tube. The high-purity elemental Ca pieces were mixed with FeAs in a ratio of 1:4 before placed into an alumina crucible, and then sealed in an evacuated quartz tube. The ampoules were heated at 500 $^\circ$C and 700 $^\circ$C for 5 h respectively, then slowly heated to 1150 $^\circ$C in 10 h, dwelled for 30 h and then cooled down to 850 $^\circ$C slowly at a rate of 1.5 $^\circ$C/h. At this temperature, some of the ampoules were taken out of furnace and quenched in ice-water rapidly, and the resulting plate-like crystals were referred to as "quenched" samples. And the samples with furnace cooling after turning off the power at 850 $^\circ$C were referred to as "furnace cooled" samples. Some of the "quenched" crystals of CaFe$_2$As$_2$ were selected for post-annealing at various temperatures for 20 h, which were called "annealed" samples. All the crystals were separated from the flux mechanically at room temperature.

All the single-crystalline samples were characterized by X-ray diffraction (XRD) analysis at room temperature, performed on a PAN-analytical diffractometer with Cu-K$_\alpha$ radiation. Temperature dependent XRD was measured by a Rigaku-TTR3 diffractometer using high-intensity graphite monochromatized Cu-K$_\alpha$ radiation. Chemical analysis was performed via energy dispersive spectroscopy (EDX), giving stoichiometry of 1 : 2 : 2 in all specimens within the instrument error range. The resistivity measurements against temperature were performed by employing a standard DC four-probe method using a Quantum Design physical property measurement system (PPMS) down to 2 K. The DC magnetic susceptibility was measured as a function of temperature with both zero field cooling (ZFC) and field cooling (FC) methods under a magnetic field of 10 Oe using a Quantum Design magnetic property measurement system (MPMS). The DC magnetic susceptibility as a function of field at 2 K was also carried out using such MPMS system to ensure the superconductivity.

## 3. Results and discussion

Fig. 1 presents the room temperature XRD patterns for the single crystals of furnace cooled, quenched and annealed samples with an image of the plate-like single crystal for the quenched specimen shown in the inset. The observation of the (00l) peaks indicates the orientation along the c-axis for all the samples. Clear changes of the lattice parameter c are revealed from the varying positions of the (008) peaks, and the c-axis parameter for all samples was calculated. The furnace cooled sample has a similar c-axis parameter to that of previously reported non-superconducting tetragonal $CaFe_2As_2$ parent compound [32]. The quenched crystal has the shortest c-axis parameter of 11.564(7) Å. After annealing processes at low temperature from 150 $^o$C to 400 $^o$C for the quenched crystal, the crystal lattice slowly relaxed and the c-axis returns to the value close to that of the furnace-cooled sample, which reveals strong internal crystallographic strain from lattice distortion and defects stabilized in the quenched crystals. This is also indicated by the varied shape of the (008) peaks, from the irregular, broad and low-intensity peaks of the quenched crystal to the typical sharp Lorentz peaks of the annealed crystal. The temperature dependent XRD patterns for a quenched crystal are shown in Fig. 2(a) from 35 K to 300 K near the (002) peaks, with the corresponding lattice parameter c plotted in Fig. 2(b). A dramatic shrinkage of the c-axis happens around 80 K, and this is the typical characteristic of the collapsed phase transition in the $CaFe_2As_2$ single crystals [23]. The stabilization of this collapsed phase by a fast quenching process is consistent with the reported phase diagram [26]. Here we note that powder XRD was not presented because grinding will cause relaxation of the crystal lattice.

The DC magnetization for all the undoped $CaFe_2As_2$ single crystals were carefully measured between 2 K and 60 K with both zero-field-cooling (ZFC) and field-cooling (FC) methods under a magnetic field of 10 Oe applied parallel to the c-axis, as shown in Fig. 3(a). No superconducting signal is observed in all the furnace cooled samples, while the quenched crystal shows a clear diamagnetic superconducting transition at the onset temperature of 25 K, as shown in the expanded curve of the inset. The isothermal magnetization curve as a function of magnetic field measured at 2 K for the quenched crystal reveals a typical characteristic of a type-II superconductor in Fig.3 (b), and the estimated superconducting shielding volume fraction is about 5%, similar to that reported in the undoped superconducting $SrFe_2As_2$ [7]. For the annealed samples, the superconducting transition quickly vanishes when the annealing temperature is above 200 $^oC$. Together with the XRD results, this suggests the strong internal lattice strain is the key for the occurrence of instable superconductivity in the undoped quenched $CaFe_2As_2$ crystal, which is slowly released after such low temperature annealing process.

The temperature dependence of normalized in-plane electrical resistivity for all the $CaFe_2As_2$ single crystals was measured from 2 K to 300 K, which is shown in Fig. 4. For the furnace cooled sample, an anomaly characteristic happens at 170 K, which is associated with the structural and AFM phase transition as happening in all the reported 122-type parent compounds of iron pnictides [6]. For the quenched crystals, the upward resistivity anomaly around 170 K is completely suppressed, and another sharp resistivity drop appears around 80 K. This is in corresponding with the happening of collapsed tetragonal phase transition as revealed by XRD measurements, which replaces the orthorhombic structural transition. We note that due to sample breakage caused by the dramatic lattice shrinkage at the collapsed transition, the resistivity data for nearly half of our quenched samples could not be measured below the transition [26]. For the samples with acquired resistivity data, a superconducting resistivity drop happens at 27 K, with the resistivity very close to zero at 2 K. This is in accord with the Meissner transition at 25 K. We note that the resistivity measurement is generally much more sensitive than magnetization measurement, and zero resistivity was reported in $BaFe_2As_2$ with no diamagnetic signal observed [8].

The reason for the absence of zero resistivity here is attributed to the formation of internal micro-cracks in the crystal during the collapsed structural transition. For the low temperature annealed crystals, no obvious change happens for the sample annealed at 150 °C. When the annealing temperature is above 200 °C, the resistivity drop at 80 K disappears, and the superconducting resistivity drop was also quickly suppressed with increasing annealing temperature. Meanwhile, the upward resistivity anomaly appears, which is the indication of the orthorhombic structural transition. For the 400 °C annealed sample, the behavior of resistivity is fully restored and similar to that of furnace cooled crystal, which is consistent with the tendency of the change in XRD measurements. Together with the magnetization data, we see that superconductivity was suppressed after the ground state changes from the collapsed tetragonal phase into the orthorhombic phase by annealing above 200 °C.

In conclusion, we found non-bulk superconductivity below 25 K in $CaFe_2As_2$ single crystals quenched from high temperature. Interestingly, the superconductivity exists in the collapsed tetragonal phase of the undoped $CaFe_2As_2$ compound at ambient pressure, which is not observed previously. The obvious shrinkage of the c-axis for the quenched crystals and the vanishing of superconductivity after low temperature annealing treatment indicate that superconductivity is induced by the strong internal crystallographic strain stabilized by the fast quenching process.

**Acknowledgments**

The authors are grateful for the financial supports from the National Natural Science Foundation of China (No. 11474339), the National Basic Research Program of China (973 Program, No. 2010CB923000 and 2011CBA00100) and the Strategic Priority Research Program of the Chinese Academy of Sciences (No. XDB07020100).

## References:

[1] KAMIHARA Y., WATANABE T., HIRANO M. and HOSONO H., *J. Am. Chem. Soc.,* **130** (2008) 3296.

[2] REN Z. A., CHE G. C., DONG X. L., YANG J., LU W., YI W., SHEN X. L., LI Z. C., SUN L. L., ZHOU F. and ZHAO Z. X., *Europhys. Lett.,* **83** (2008) 17002.

[3] DE LA CRUZ C., HUANG Q., LYNN J. W., LI J., RATCLIFF W. II, ZARESTKY J. L., MOOK H. A., CHEN G. F., LUO J. L., WANG N. L. and DAI P.C., *Nature*, **453** (2008) 899.

[4] SEFAT A. S., JIN R., MCGUIRE M. A., SALES B. C., SINGH D. J. and MANDRUS D., *Phys. Rev. Lett.,* **101** (2008) 117004.

[5] GAO Z., QI Y., WANG L., WANG D., ZHANG X., YAO C., WANG C. and MA Y., *Europhys. Lett.,* **95** (2011) 67002.

[6] KASINATHAN D., ORMECI A., KOCH K., BURKHARDT U., SCHNELLE W., LEITHE-JASPER A. and ROSNER H., *New J. Phys.,* **11** (2009) 025023.

[7] SAHA S. R., BUTCH N. P., KIRSHENBAUM K., PAGLIONE J. and ZAVALIJ P. Y., *Phys. Rev. Lett.,* **103** (2009) 037005.

[8] KIM J. S., BLASIUS T. D., KIM E. G. and STEWART G. R., *J. Phys.: Condens. Matter* **21** (2009) 342201.

[9] TORIKACHVILI M. S., BUD'KO S. L., NI N. and CANFIELD P. C., *Phys. Rev. Lett.,* **101** (2008) 057006.

[10] PARK T., PARK E., LEE H., KLIMCZUK T., BAUER E. D., RONNING F. and THOMPSON J. D., *J. Phys.: Condens. Matter* **20** (2008) 322204.

[11] KREYSSIG A., GREEN M. A., LEE Y., SAMOLYUK G. D., ZAJDEL P., LYNN J. W., BUD'KO S. L., TORIKACHVILI M. S., NI N., NANDI S., LEÃO J. B., POULTON S. J., ARGYRIOU D. N., HARMON B. N., MCQUEENEY R. J., CANFIELD P. C. and GOLDMAN A. I., *Phys. Rev. B*, **78** (2008) 184517.

[12] GOLDMAN A. I., KREYSSIG A., PROKEŠ K., PRATT D. K., ARGYRIOU D. N., LYNN J. W., NANDI S., KIMBER S. A. J., CHEN Y., LEE Y. B., SAMOLYUK G., LEÃO J. B., POULTON S. J., BUD'KO S. L., NI N., CANFIELD P. C., HARMON B. N. and MCQUEENEY R. J., *Phys. Rev. B*, **79** (2009) 024513.

[13] YU W., ACZEL A. A., WILLIAMS T. J., BUD'KO S. L., NI N., CANFIELD P. C. and LUKE G. M., *Phys. Rev. B*, **79** (2009) 020511(R).

[14] CANFIELD P. C., BUD'KO S. L., NI N., KREYSSIG A., GOLDMAN A. I., MCQUEENEY R. J., TORIKACHVILI M. S., ARGYRIOU D. N., LUKE G. and YU W., *Physica C* **469** (2009) 404.

[15] LEE H., PARK E., PARK T., SIDOROV V. A., RONNING F., BAUER E. D. and THOMPSON J. D., *Phys. Rev. B*, **80** (2009) 024519.

[16] LV B., DENG L., GOOCH M., WEI F., SUN Y., MEEN J. K., XUE Y. Y., LORENZ B. and CHU C. W., *Proc. Natl. Acad. Sci. USA*, **108** (2011) 15705.

[17] DENG L. Z., LV B., ZHAO K., WEI F. Y., XUE Y. Y., WU Z. and CHU C. W., *Phys. Rev. B*, **93** (2016) 054513.

[18] NI N., NANDI S., KREYSSIG A., GOLDMAN A. I., MUN E. D., BUD'KO S. L. and CANFIELD P. C., *Phys. Rev. B*, **78** (2008) 014523.

[19] RONNING F., KLIMCZUK T., BAUER E. D., VOLZ H. and THOMPSON J. D., *J. Phys.:*

*Condens. Matter* **20** (2008) 322201.
[20] CHOI K. Y., WULFERDING D., LEMMENS P., NI N., BUD'KO S. L. and CANFIELD P. C., *Phys. Rev. B*, **78** (2008) 212503.
[21] GOLDMAN A. I., ARGYRIOU D. N., OULADDIAF B., CHATTERJI T., KREYSSIG A., NANDI S., NI N., BUD'KO S. L., CANFIELD P. C. and MCQUEENEY R. J., *Phys. Rev. B*, **78** (2008) 100506(R).
[22] PRATT D. K., ZHAO Y., KIMBER S. A. J., HIESS A., ARGYRIOU D. N., BROHOLM C., KREYSSIG A., NANDI S., BUD'KO S. L., NI N., CANFIELD P. C., MCQUEENEY R. J. and GOLDMAN A. I., *Phys. Rev. B*, **79** (2009) 060510(R).
[23] YILDIRIM T., *Phys. Rev. Lett.,* **102** (2009) 037003.
[24] ORTENZI L., GRETARSSON H., KASAHARA S., MATSUDA Y., SHIBAUCHI T., FINKELSTEIN K. D., WU W., JULIAN S. R., KIM Y. J., MAZIN I. I. and BOERI L., *Phys. Rev. Lett.,* **114** (2015) 047001.
[25] BUD'KO S. L., MA X., TOMIĆ M., RAN S., VALENTÍ R. and CANFIELD P. C., *Phys. Rev. B*, **93** (2016) 024516.
[26] RAN S., BUD'KO S. L., PRATT D. K., KREYSSIG A., KIM M. G., KRAMER M. J., RYAN D. H., ROWAN-WEETALUKTUK W. N., FURUKAWA Y., ROY B., GOLDMAN A. I. and CANFIELD P. C., *Phys. Rev. B*, **83** (2011) 144517.
[27] SAPAROV B., CANTONI C., PAN M., HOGAN T. C., RATCLIFF W. II, WILSON S. D., FRITSCH K., GAULIN B. D., SEFAT A. S. and TACHIBANA M., *Sci. Rep.*, **4** (2014) 4120.
[28] ZHAO K., STINGL C., MANNA R. S., JIN C. Q. and GEGENWART P., *Phys. Rev. B*, **92** (2015) 235132.
[29] KASAHARA S., SHIBAUCHI T., HASHIMOTO K., NAKAI Y., IKEDA H., TERASHIMA T. and MATSUDA Y., *Phys. Rev. B*, **83** (2011) 060505(R).
[30] DANURA M., KUDO K., OSHIRO Y., ARAKI S., C. KOBAYASHI T. and NOHARA M., *J. Phys. Soc. Jpn.,* **80** (2011) 103701.
[31] SAHA S. R., BUTCH N. P., DRYE T., MAGILL J., ZIEMAK S., KIRSHENBAUM K., ZAVALIJ P. Y., LYNN J. W. and PAGLIONE J., *Phys. Rev. B*, **85** (2012) 024525.
[32] WU G., CHEN H., WU T., XIE Y. L., YAN Y. J., LIU R. H., WANG X. F., YING J. J. and CHEN X. H., *J. Phys.: Condens. Matter* **20** (2008) 422201.

**Figure Captions:**

Figure 1: The room temperature XRD patterns for the $CaFe_2As_2$ single crystals of furnace cooled, quenched and annealed samples with the (008) reflection expanded in the right panel. The inset shows an image of a plate-like single crystal for the quenched sample.

Figure 2: (a) The XRD patterns around (002) reflection for a quenched crystal with temperature from 35 K to 300 K. (b) The temperature dependence of lattice parameter $c$ for the quenched crystal.

Figure 3: (a) The temperature dependence of magnetic susceptibility for all the $CaFe_2As_2$ single crystals between 2 K and 60 K with both ZFC and FC measurements. (b) The isothermal magnetization curve at 2 K for the quenched crystal.

Figure 4: The temperature dependence of normalized in-plane electrical resistivity for the $CaFe_2As_2$ single crystals of (a) furnace cooled sample, (b) four quenched samples, and (c) annealed samples. The insets in (b) show the expanded resistivity curves for the quenched samples around the collapsed phase transition and the superconducting phase transition respectively. Some samples like 'S4' cannot be measured for resistivity below the collapsed phase transition temperature due to sample breakage.

**Fig. 1.**

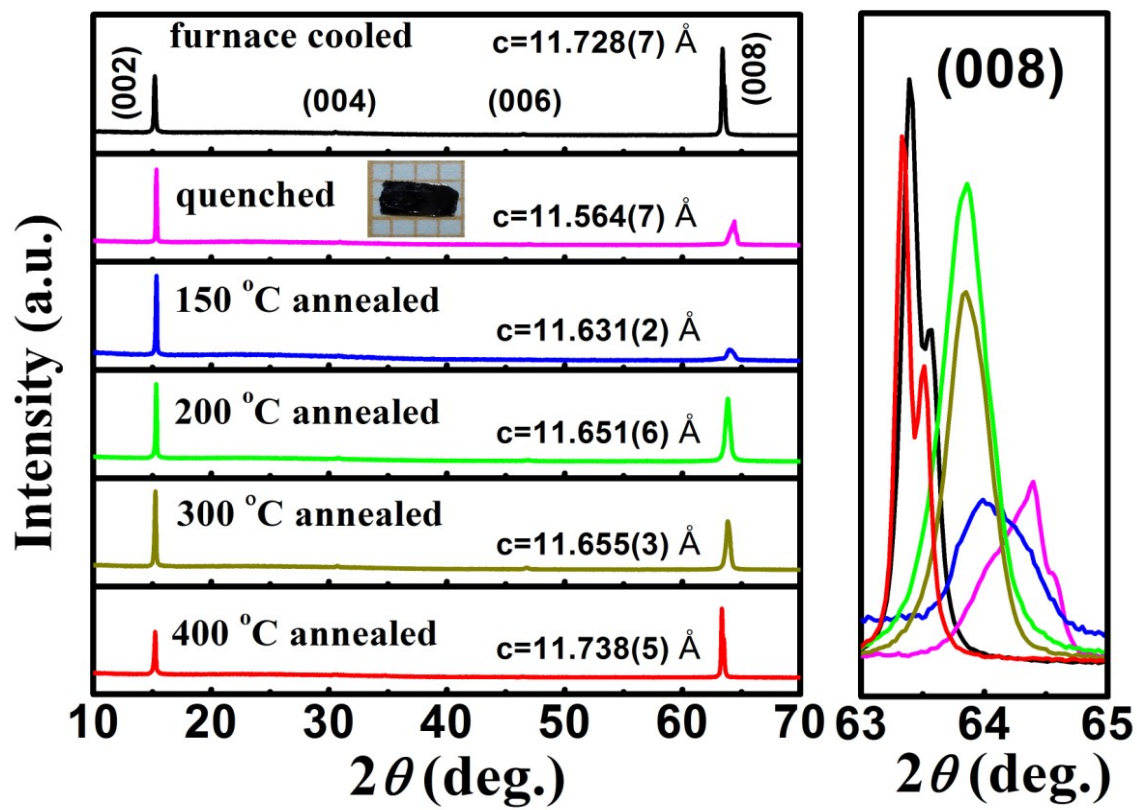

**Fig. 2.**

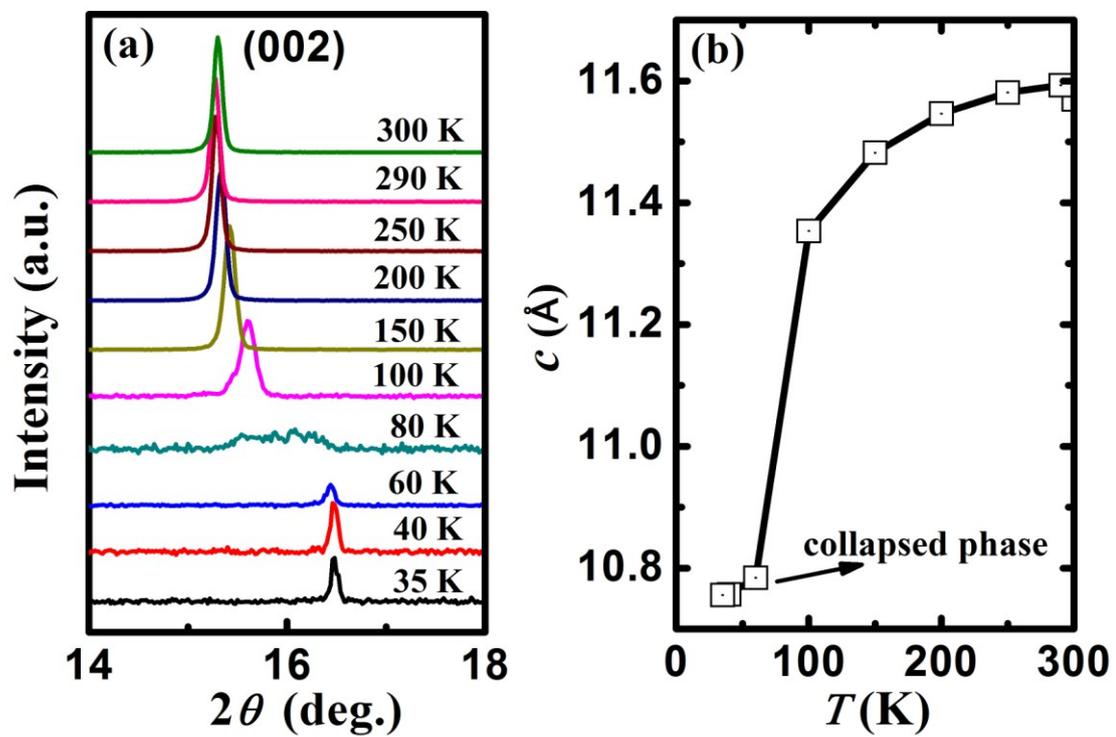

**Fig. 3.**

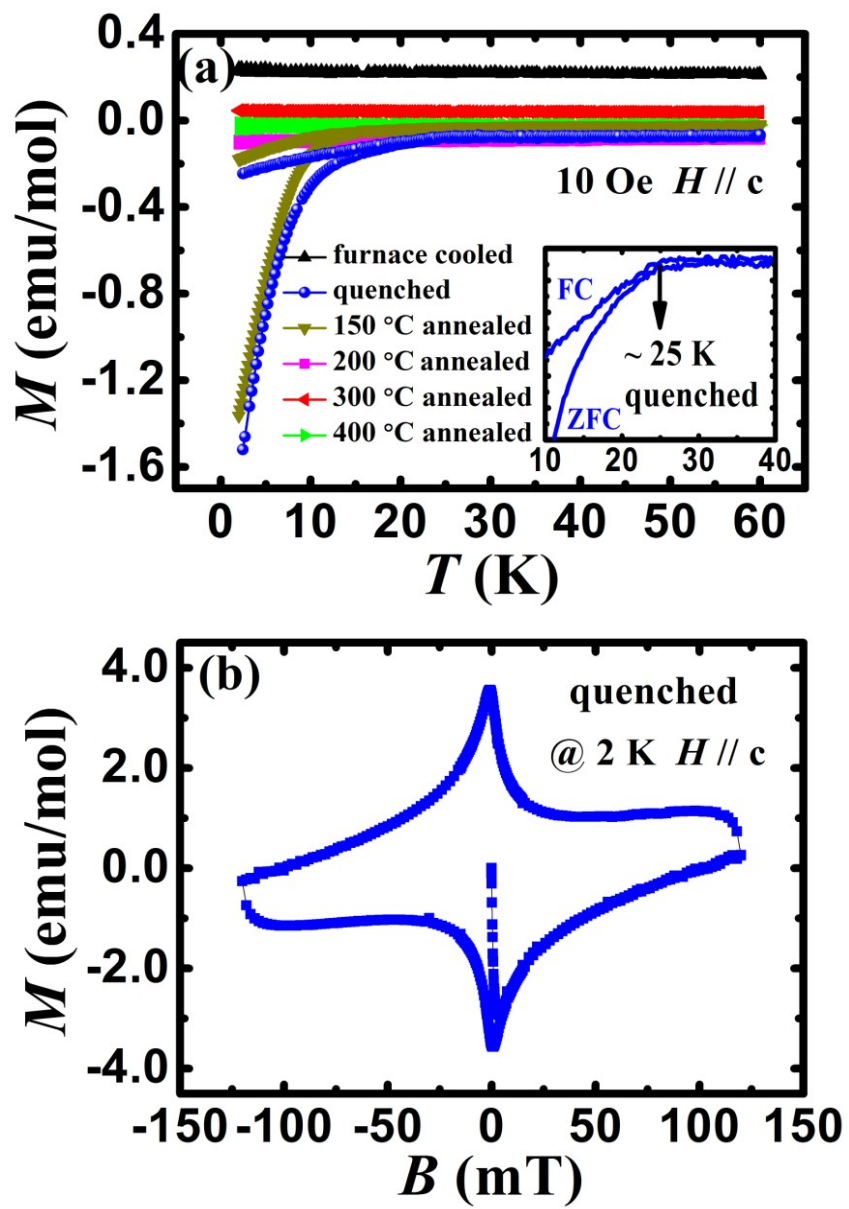

**Fig. 4.**

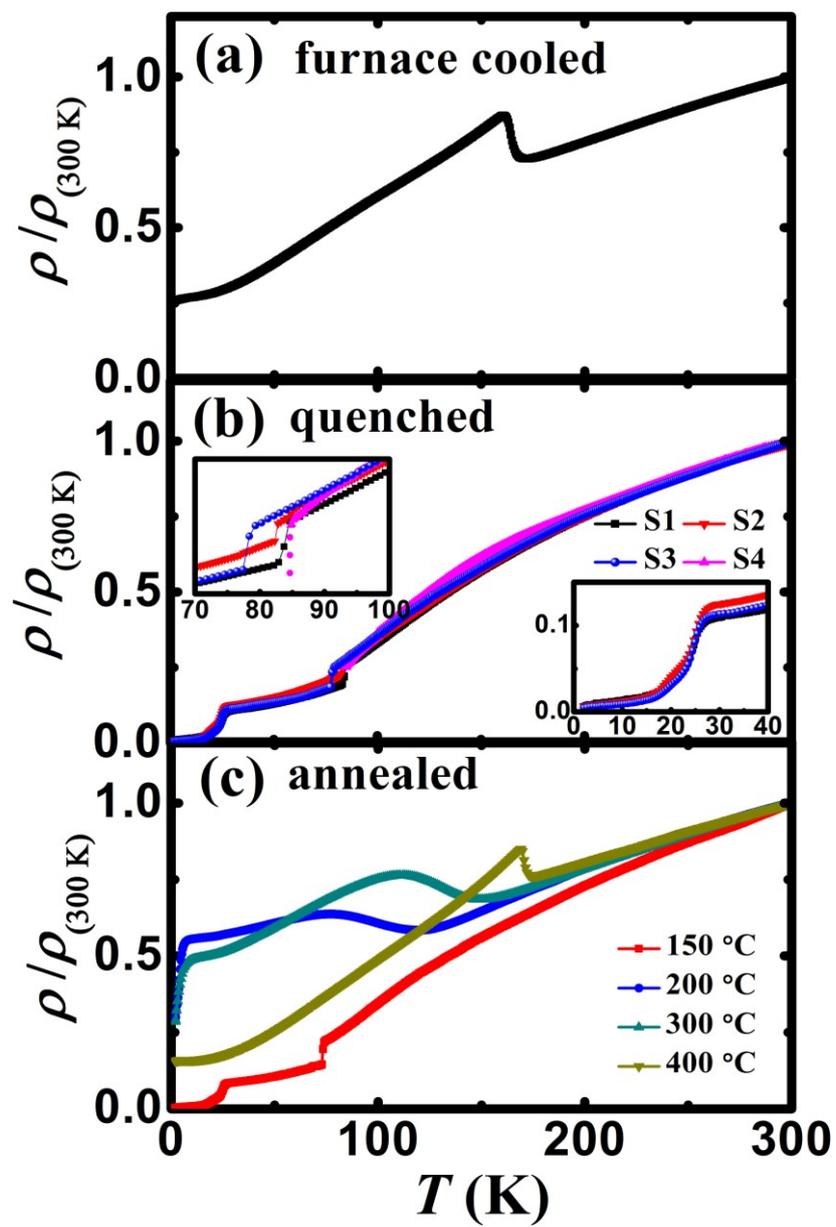